# The Use of Self Organizing Map Method and Feature Selection in Image Database Classification System

**Dian Pratiwi[1]**

[1] **Department of Information Engineering, Trisakti University
Jakarta, 15000, Indonesia
pratiwi.dian@gmail.com**

**Abstract**
This paper presents a technique in classifying the images into a number of classes or clusters desired by means of Self Organizing Map (SOM) Artificial Neural Network method.
A number of 250 color images to be classified as previously done some processing, such as RGB to grayscale color conversion, color histogram, feature vector selection, and then classifying by the SOM Feature vector selection in this paper will use two methods, namely by PCA (Principal Component Analysis) and LSA (Latent Semantic Analysis) in which each of these methods would have taken the characteristic vector of 50, 100, and 150 from 256 initial feature vector into the process of color histogram. Then the selection will be processed into the SOM network to be classified into five classes using a learning rate of 0.5 and calculated accuracy.
Classification of some of the test results showed that the highest percentage of accuracy obtained when using PCA and the selection of 100 feature vector that is equal to 88%, compared to when using LSA selection that only 74%. Thus it can be concluded that the method fits the PCA feature selection methods are applied in conjunction with SOM and has an accuracy rate better than the LSA feature selection methods.

## 1. Introduction

Image classification technology is currently being widely developed. This is because such techniques can support and facilitate image retrieval or CBIR (Content Based Image Retrieval) which desired for some image data in a large scale.
The image meant in this paper is a collection of natural images (total 250 images) that contain ordinary objects such as mountains, rivers, cars, animals, and flowers which will then be grouped into different classes by means of SOM artificial neural network methods.
The SOM (Self Organizing Map) Neural Network or commonly called a Kohonen Neural Network system is one of the unsupervised learning model that will classify the units by the similarity of a particular pattern to the area in the same class.

In the presence of this classification technique is also expected to speed up the search image data needed, because the search did not match the image again with a whole one by one in the database but starting from image to the class corresponding to the query image are included

## 2. Feature Vector Representation

The image will be grouped into clusters previously required two prior processing, the color conversion and histogram to produce a feature vector of a set of image.

2.1 Color Conversion

Color conversion is referred to in this paper is the conversion of RGB color images (24 bits) into grayscale (8 bits), so that the color model will be simpler with each pixel grey level between 0 to 255. The conversion formula is :

$$RGB = \frac{R+G+B}{3} \quad (1)$$

Where $R$ is the red value of pixel, $G$ is the green value and $B$ is the blue value of pixel images.

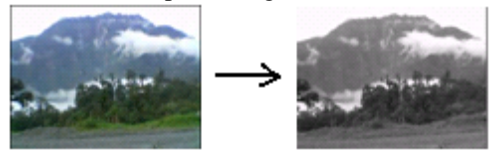

Fig. 1  RGB Color Conversion into Grayscale.

2.2 Histogram

Histogram or color histogram is one of the techniques of statistical features that can be used to take the feature vector of a data set, images, video or text.
The generated feature vector of color histogram will be a probability value of $h_i$, $n_i$ is the number of pixels of $i$ color intensity that appears in the $m$ image divided by the total $n$ image pixel.

$$h_i = \frac{n_i}{n} \quad i = 0 \ldots 255 \tag{2}$$

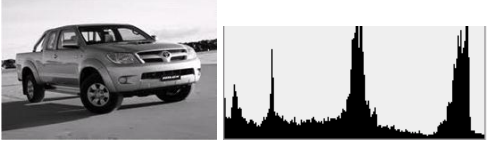

Fig. 2  Sample Grayscale Image and Color Histogram.

In this process feature vector matrix obtained from the whole image will be sized ($i$ x $m$), which will then be carried through the characteristic dimension reduction methods of feature selection on the next process.

## 3. Feature Vector Selection

Feature vector selection technique in this paper focuses on two methods namely PCA (Principal Component Analysis) and LSA (Latent Semantic Analysis), which is a pre-processing stage of the classifying process.
The purpose of this feature vector selection is to reduce the K dimensional matrix obtained from the characteristic features of the previous vector to < K dimensions without reducing the important information on it [9].

### 3.1 Principal Component Analysis (PCA)

PCA method is a global feature selection algorithm which first proposed by Hotteling (1933) as a way to reduce the dimension of a space that is represented in statistics of variables ($x_i$, $i = 1,2\ldots n$) which mutually correlated with each other [5].
During its development, PCA algorithm (also called the Hotteling transformation) can be used to reduce noise and extract features or essential characteristics of data before the classification process.
Election of global feature selection technique in this case because the images will be classified to have relatively low frequency (low level), so the PCA method is quite suitable to be applied.
The steps in the PCA algorithm namely [6]:
  a) Create a matrix [$X_1$, $X_2$, .... $X_m$] which representing $N^2$ x $m$ data matrix. $X_i$ is the image of size $N$ x $N$, where $N^2$ is the total pixels of the image dimensions and $m$ is the number of images to be classified.
  b) Use the following equation to calculate the average value of all images :
$$Y = \frac{1}{m}\sum_{i=1}^{m} X_i \tag{3}$$
  c) Calculated the difference matrix :
$$\overline{X}_i = X_i - Y \tag{4}$$
  d) Use the difference matrix obtained previously to generated the covariance matrix to obtain the correlation matrix :
$$\Sigma = \sum_{i=1}^{N} \overline{x}_i \overline{x}_i^T \tag{5}$$
  e) Use the correlation matrix to evaluate the eigenvector :
$$\Sigma \phi_i = \lambda \phi_i \tag{6}$$
  Where $\phi$ is orthogonal eigenvector matrix, $\lambda$ is the eigenvalue diagonal matrix with diagonal elements sorted ($\lambda_0 > \lambda_1 \ldots > \lambda_N^{2-1}$ and $\lambda_0 = \lambda_{max}$), which aims to reduce the eigenvector matrix form using the feature space $\Phi$.
  The order of the eigenvectors with the largest eigenvalue represents the data closer or similar to the original data [2].
$$\Phi = [\phi_1 | \phi_2 | \ldots | \phi_n] \tag{7}$$
  Where, $1 \leq n \leq N^2$.
  f) If $\Phi$ is a feature vector of the sample image $X$, then :
$$y_n = \Phi^T \overline{X}_i \tag{8}$$
  With feature vector $y$ is the $n$-dimensional.

### 3.2 Latent Semantic Analysis (LSA)

LSA is a statistical method that was originally used in the field of Artificial Intelligence branch of Natural Language Processing (NLP) to analyze the data as plain text or document.
LSA or also known as LSI (Latent Semantic Indexing), in its application can also be used to process image data to form a new matrix decomposition of the initial matrix into three matrices which correlated. This technique is then called by the Singular Value Decomposition (SVD), which is part of the LSA [7].
$$X = USV^T \tag{9}$$
If $X$ is a matrix of feature vector are sized $i$ x $m$, with total of grey level $i$ and the total image $m$, then $U$ is a matrix of othonormal $i$ x $m$, $S$ is the $m$ x $m$ diagonal matrix, and $V$ are orthonormal $m$ x $m$.
The matrix $U$ obtained from the search eigenvectors of the matrix multiplication yields $X.V.S^{-1}$ [8]. While the $V$ matrix obtained from the search eigenvectors of the matrix multiplication yields $X^T.X$.
The $S$ Diagonal matrix will contain the eigenvalues of the matrix of $V^T$ orthonormal eigenvectors with a sequence starting from the largest eigenvalue to the smallest, from left to right.
After the third matrix are obtained, then the reduction in number of $r$-dimensional matrix as a vector which will produce the $U_k$, $S_k$ and $V_k^T$ matrix. Then to generate the $Q$

feature vector matrix with K or n-dimensional, use the formula :

$$Q = Q^T . U_k . S_k^{-1} \qquad (10)$$

Where $S_k^{-1}$ is a diagonal matrix inverse, and $Q^T$ is a new feature vector transpose matrix.

## 4. Self Organizing Map (SOM)

Artificial Neural Network (ANN) is defined as an information processing system that has characteristics resembling human neural tissue. The existence of ANN provides a new technology to help solve problems that require thinking of experts and computer based routine.

A few of ANN application was for classification system (clustering), association, pattern recognition, forecasting and analysis [1][3]. And in this paper, the ANN method which applied is SOM method, that to be used to classify the image into a set of five different classes.

In the Self Organizing Map (SOM) method, the applied learning is an unsupervised learning where the network does not utilize the class membership of sample training, but use the information in a group of neurons to modify the local parameter [3].

The SOM system is adaptively classify samples (X image data) into classes determined by selecting the winning neurons are competitive and the weights are modified.

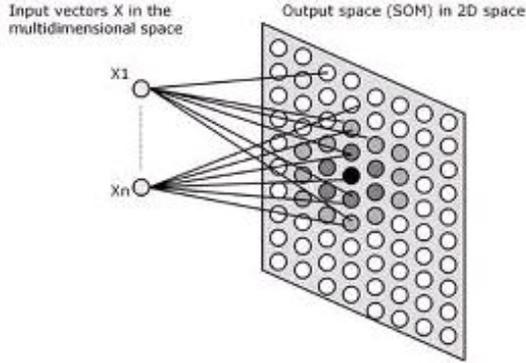

Fig. 3  SOM Artificial Neural Network.

The algorithm on the SOM neural network as follows [4] :
a. If the feature vector matrix of size $k$ x $m$ ($k$ is the number of feature vector dimensions, and $m$ is the number of data), the initialization :
   - The number of the desired $j$ class or cluster
   - The number of component $i$ of the feature vector matrix ($k$ is the row of matrix)
   - The number of vector $X_{m,i}$ = amount of data(matrix column)
   - The initial weights $W_{ji}$ were randomly with interval 0 to 1
   - The initial learning rate $\alpha(0)$
   - The number of iteration ($e$ epoch)
b. Execute the first iteration until the total iteration (epoch)
c. Calculate the vector image to start from 1 to $m$ :

$$D(j) = \sum_i (W_{ji} - X_{m,i})^2 \qquad (11)$$

   For all of $j$

   - Then determine the minimum value of $D(j)$
   - Make changes to the $j$ weight with the minimum of $D(j)$

$$W_{ji}^{new} = W_{ji} + \alpha ( X_{m,i} - W_{ji} ) \qquad (12)$$

d. Modify the learning rate for the next iteration :

$$\alpha(t + 1) = 0,5\ \alpha(t) \qquad (13)$$

   which $t$ start from the first iteration to $e$.

e. Test the termination condition
   Iteration is stopped if the difference between $W_{ji}$ and $W_{ji}$ the previous iteration only a little or a change in weights just very small changes, then the iteration has reached convergence so that it can be stopped.

f. Use a weight of $W_{ji}$ that has been convergence to grouping feature vector for each image, by calculating the distance vector with optimal weights.

g. Divide the image ($X_m$) into classes :
   If $D(1)<D(2)<D(3)<D(4)<D(5)$, then the images included in class 1

   If $D(2)<D(1)<D(3)<D(4)<D(5)$, then the images included in class 2

   If $D(3)<D(2)<D(1)<D(4)<D(5)$, then the images included in class 3

   If $D(4)<D(3)<D(2)<D(1)<D(5)$, then the images included in class 4

   If $D(5)<D(4)<D(3)<D(2)<D(1)$, then the images included in class 5

## 5. Implementation

In this paper, the implementation stage of image classification system with SOM and feature selection method can be viewed via the following flowchart :

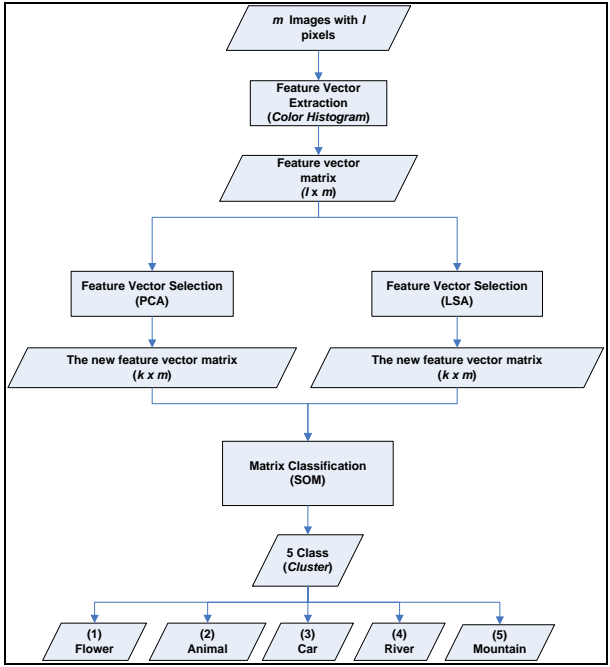

Fig. 4  Flowchart of The Implementation

Figure 4 shows that some processing needs to be done previously until the images can be classified into five classes, which in this study consists of five classes namely flower class, animal class, car class, river class, and mountain classes.

The image used in this study were as many as 250 images by dividing each image of the 50 pieces in each class.

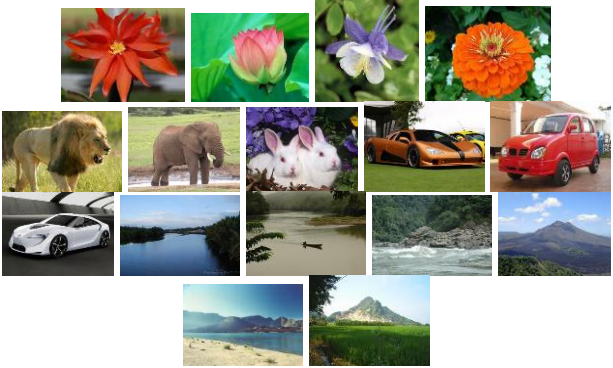

Fig. 5  Example of 16 Images to be Classified

In the color histogram, feature vector of 250 images is retrieved and will generate feature vector matrix $i$ x $m$, which is 256 x 250. 256 is the number of grey level (0 – 255) from the grayscale color conversion process.

The matrix will then be reduced to the dimensions as $r$ columns or rows through PCA and LSA feature selection algorithms. From the result of algorithm, will be gained a new feature matrix to the size $k$ x $m$ ($k$ is the dimension after reduction, and $m$ is the total images).

So if the initial dimensions of the image matrix is 256x250, then after feature selection (eg, $r$ reduction as much as 156 vectors), the size of the matrix will be changed to 100x250.

Then from the matrix $k$ x $m$ will be classified into five classes by SOM neural network algorithm, with the network parameters that are used :

- The number of class $(j) = 5$
- The number of vector component $(i) = 50 ; 100 ; 150$
- The number of $X$ vectors $= 250$
- The initial weights $(W_{ji}) = 0$ to 1 (random)
- The initial learning rate $(\alpha) = 0.5$
- Total iteration or epoch $(e) = 500$

The experimental results will be seen from the large percentage of SOM accuracy in classifying the images corresponding to the class by using an application program, where the formula of accuracy is following :

$$\% = \frac{\text{The image classification results}}{\text{The number of image in the class should be}} \times 100\% \qquad (14)$$

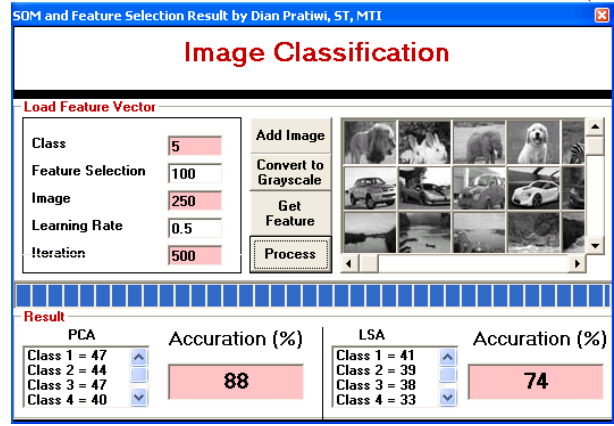

Fig. 6  The Display of Image Classification Program with SOM Method and Feature Selection

From the total tests performed in this study, the results of image classification can be demonstrated by the following tables :

Table 1: Classification Results of PCA and SOM with The 100 Feature Vectors

| Classification Result | | | | | The class |
| Flower | Animal | Car | River | Mountain | should be |
|---|---|---|---|---|---|
| 47 | 1 | 2 | 0 | 0 | Flower |
| 3 | 44 | 1 | 1 | 1 | Animal |
| 0 | 2 | 47 | 0 | 1 | Car |
| 1 | 3 | 1 | 40 | 5 | River |
| 1 | 1 | 0 | 4 | 44 | Mountain |

Table 2: Classification Results of SOM with The LSA and PCA Method for Feature Vectors as many as 50, 100, and 150 Vectors

|  | PCA (Feature Vector) | | | LSA (Feature Vector) | | |
| --- | --- | --- | --- | --- | --- | --- |
|  | 50 | 100 | 150 | 50 | 100 | 150 |
| Flower | 42 | 47 | 43 | 40 | 41 | 46 |
| Animal | 41 | 44 | 41 | 33 | 39 | 35 |
| Car | 39 | 47 | 42 | 34 | 38 | 42 |
| River | 29 | 40 | 35 | 22 | 33 | 25 |
| Mountain | 27 | 44 | 39 | 24 | 35 | 30 |
| ∑ | 178 | 222 | 200 | 153 | 186 | 178 |
| % Accuration | 71 | 88 | 80 | 61 | 74 | 71 |

In general, it can be seen from Table 2 that the SOM classification uses of both PCA and LSA feature selection showed a fairly good percentage of accuracy, with an average success percentage of 68.6% - 79.6%. But when compared to results from the use of both feature selection, the highest level of accuracy obtained when using the PCA by 88% (222 of 250 classified image is the right image) with the number of feature vector is 100 vector (shown in Table 1). While the LSA, the highest accuracy results obtained only by 74% (186 of 250 classified image is the right image) with the same number of feature vectors (shown in Figure 6).

Thus, it can be concluded that the SOM method is more suitable to be applied with PCA feature selection to classify images into classes in accordance with good results compared with LSA feature selection.

# 6. Conclusions

The conclusion to be drawn from the writing of this paper are :
1. The selection method of PCA and LSA feature selection precise enough to be implemented in the image classification system, because it can reduce the dimensions of the image matrix while producing a high level of accuracy which is between 61% - 88%
2. The ability of SOM classification techniques can be further increased when the number of image feature vector as many as 100 vector with the type of feature selection is PCA. This is because the number of feature vectors is a number of feature vector with the best accuracy, that is 88%, of which there remain some important information in these vectors although the dimension of characteristic matrix has been reduced.
3. With the image classification techniques in the database will be able to facilitate and speed up image retrieval system, because the image can look directly into the appropriate classes without having to search one by one from each class.

# References

[1] Siong, A.W, *Pengenalan Citra Objek Sederhana dengan Menggunakan Metode Jaringan Syaraf Tiruan SOM*. *Prosiding Seminar Nasional I Kecerdasan Komputasional*, Universitas Indonesia. 1999.
[2] H. C. Wu, S. Huang, "User Behavior Analysis in Masquerade Detection Using Principal Component Analysis". *Eighth International Conference on Intelligent Systems Design and Application (ISDA)*, 201-206, DOI : 10.1109/ISDA.2008.243.
[3] A. Hermawan, *Jaringan Syaraf Tiruan Teori dan Aplikasi*. Yogyakarta : Penerbit ANDI, 2006.
[4] J.J. Siang, *Jaringan Syaraf Tiruan dan Pemrogramannya menggunakan Matlab*, Yogyakarta : Penerbit Andi, 2005
[5] Jolliffe, I.T., *Principal Component Analysis*, 2$^{nd}$ ed., New York : Springer-Verlag, 2002
[6] C. J. Lin, C.H. Chu, C.Y. Lee, Y.T. Huang, "2D/3D Face Recognition Using Neural Networks Based on Hybrid Taguchi Particle Swarm Optimization", *Eighth International Conference on Intelligent Systems Design and Application (ISDA)*, 307-312, DOI : 10.1109/ISDA.2008.286.
[7] T.K. Laundauer, P.W. Foltz, D. Laham, *Introduction to Latent Semantic Analysis*, Discourse Process, 25, 259-284, 1998.
[8] M.E. Wall, A. Rechtsteiner, L.M. Rocha, *Singular Value Decomposition and Principal Component Analysis*, Portland State University, 2003.
[9] N. Navaroli, D. Turner, A.I. Conception, R.S. Lynch, "Performance Comparison of ADRS and PCA as a Preprocessor to ANN for Data Mining", *Eighth International Conference on Intelligent Systems Design and Application (ISDA)*, 47-52, DOI : 10.1109/ISDA.2008.133

**Dian Pratiwi (26)** was born in Jakarta – Indonesia on February 25, 1986. The last education is The Master of Information Technology achieved in 2011 at The University of Bina Nusantara, Jakarta – Indonesia with a GPA of 3,66. Her bachelor achieved at Trisakti University, Jakarta – Indonesia with the title of "Very Satisfied" and was awarded as "One of The 7 Best Graduate Department of Information Engineering" in 2007.

Now, the author worked as a lecturer at Trisakti University also taught courses in image processing, mobile programming, web based programming, and computer graphics began in 2008 until now. Some writings ever made that has been published is entitled "An Application of Backpropagation Artificial Neural Network Method for Measuring The Severity of Osteoarthritis", which in 2011 published in the journal IJENS – IJET. In addition, the author also wrote another article which successfully published nationally in Indonesia at the SNTI seminar : (Jakarta, Indonesia : Trisakti University, 2010) with the title "Sistem Deteksi Penyakit Pengeroposan Tulang dengan Metode Jaringan Syaraf Tiruan Backpropagation dan Representasi Ciri dalam Ruang Eigen" and SITIA seminar (Surabaya, Indonesia : Institute of Ten November Technology – ITS, 2011) entitled "Penerapan Metode Jaringan Syaraf Tiruan Backpropagation dalam Mengukur Tingkat Keparahan Penyakit Osteoarthritis" which results in the proceedings. Now, she keep trying to develop her research on the medical by trying to apply her research interests are in Artificial Intelligence, digital image processing, and data mining.